# Bichromatic Moiré Superlattices for Tunable Quadrupolar Trions and Correlated States

**Authors**: Mingfeng Chen[1,*], Runtong Li[1,*], Haonan Wang[1,*], Yuliang Yang[1], Yiyang Lai[1], Chaowei Hu[2], Takashi Taniguchi[3], Kenji Watanabe[4], Jiaqiang Yan[5], Jiun-Haw Chu[2], Erik Henriksen[1,6,7], Chuanwei Zhang[1,6,7], Li Yang[1,6,7], and Xi Wang[1,6,7,#]

[1]Department of Physics, Washington University, St. Louis, MO 63130, USA
[2]Department of Physics, University of Washington, Seattle, WA, USA
[3]Research Center for Materials Nanoarchitectonics, National Institute for Materials Science, 1-1 Namiki, Tsukuba 305-0044, Japan
[4]Research Center for Electronic and Optical Materials, National Institute for Materials Science, 1-1 Namiki, Tsukuba 305-0044, Japan
[5]Materials Science and Technology Division, Oak Ridge National Laboratory, Oak Ridge, Tennessee, 37831, USA
[6]Center for Quantum Leaps, Washington University, St. Louis, MO 63130, USA
[7]Institute of Materials Science and Engineering, Washington University, St. Louis, MO 63130, USA

[*]These authors contribute equally to this work.
[#]Correspondence to: wxi@wustl.edu

## Abstract

Moiré superlattices in transition metal dichalcogenide heterostructures provide a platform to engineer many-body interactions. Here, we realize a bichromatic moiré superlattice in an asymmetric $WSe_2/WS_2/WSe_2$ heterotrilayer by combining R- and H-stacked bilayers with mismatched moiré wavelengths. This structure hosts fermionic quadrupolar moiré trions—interlayer excitons bound to an opposite-layer hole—with vanishing dipole moments. These trions arise from hybridized moiré potentials enabling multiple excitonic orbitals with tunable interlayer coupling, allowing control of excitonic and electronic ground states. We show that an out-of-plane electric field could effectively reshape moiré excitons and interlayer-intralayer electron correlations, driving a transition from interlayer to intralayer Mott states with enhanced Coulomb repulsion. The asymmetric stacking further enriches excitonic selection rules, broadening opportunities for spin-photon engineering. Our results demonstrate bichromatic moiré superlattices as a reconfigurable platform for emergent quantum states, where quadrupolar moiré trion emission may enable coherent and entangled quantum light manipulation.

## Introduction

When two or more 2D layers in a heterostructure have slightly different lattice constants or are oriented in a way that their periodicities do not perfectly align, they create a moiré pattern, i.e., moiré superlattice. It dictates various physical properties such as the electronic band structures and excitonic states et al[1–7]. Moiré superlattices demonstrate significant potentials for studying many-body interactions, including excitonic complexes (e.g., trions, biexcitons)[8–10], for quantum simulations such as Hubbard model physics[11–15]. The creation of moiré superlattices in transition metal dichalcogenide (TMD) heterostructures has revealed or predicted a diverse range of phases in these systems, including generalized Wigner crystals[16–23], Mott insulators[12,18,24], bosonic (excitonic) insulators[20,25–29], charge transfer insulators[30], as well as fractional[31–34] and integer[35] quantum anomalous states.

While enhancing the versatility of superlattice landscapes opens up a spectrum of possibilities, the inherent rigidity in these heterostructural moiré superlattices presents a significant challenge for post-fabrication tuning. One promising method is to harness the interference of distinct moiré wavelengths to craft intricate, tunable bichromatic patterns[36]. While theoretical insights have demonstrated the potential of creating adjustable multi-chromatic superlattices within TMD heterotrilayers to leverage moiré potentials from independently controlled heterointerfaces across layered structures, the experimental realization remains absent. Key open questions include how different moiré potentials interact and coevolve, how charge redistributing and stacking order vary under distinct moiré periodicities, and what role quasiparticle dynamics play in such synergistic superlattices.

Here, we report engineered heterotrilayer combining an R-stacked top bilayer and an H-stacked bottom bilayer with mismatched moiré wavelengths, resulting in a bichromatic moiré superlattice formed by the interference of two distinct moiré periods. This unique lattice geometry gives rise to emergent phenomena not accessible in conventional moiré systems. First, the hybridized moiré potentials support a new class of charged excitons—fermionic quadrupolar moiré trions—which are distinct from previously reported bosonic quadrupolar excitons. These trions exhibit electrically tunable dipole moments, enabling controlled studies of quasiparticle dynamics in complex potential landscapes. Second, bichromatic interference enables in-situ tunability of the beating periodicity, dynamically reshaping the carrier-trapping landscape. We observe an electric-field-driven transition from an interlayer Mott insulating state to an intralayer Mott state with enhanced on-site Coulomb repulsion. Further tuning modifies the moiré potential profile and shifts the density of ground-state sites, allowing precise control over electron correlations and establishing the asymmetric heterotrilayer as a reconfigurable platform for exploring programmable correlated quantum states. Finally, the coexistence of R- and H-stacked domains enables tailored control of interlayer excitonic selection rules, advancing the engineering of spin-photon coupling for quantum optoelectronic applications.

## Results

### Structures of R- and H-stacked $WSe_2/WS_2/WSe_2$ heterotrilayers

We fabricated dual-gate $WSe_2/WS_2/WSe_2$ heterotrilayer devices, which allows independent control of doping levels and vertical electric field (Fig. 1a, see Methods). The structure of heterotrilayers investigated is shown in Fig. 1b inset. We highlight the layer alignments with arrows: parallel indicating R stacking and opposite indicating H stacking. Device 1 (Fig. 1b optical image) displays the top $WSe_2$ layer aligned with the middle $WS_2$ layer at the angle close to 1° (R stacking), while the bottom WSe$_2$ layer is rotated 180° with a slightly different alignment angle (H stacking). This device also includes a bilayer region for direct comparison.

Moiré superlattices can be clearly identified in the Piezoresponse Force Microscopy (PFM) image of Device 1's trilayer region (Fig. 1c). Fourier transform data (Fig. 1c inset) reveals two distinct moiré superlattices with lattice constants approximately 7.4 nm (~0.9°) and 6.7 nm (~1.5°). Besides, a larger superlattice background could be identified in the PFM image emerging from the interference of these two sets. This small, deliberate twist-angle offset of roughly 0.5° keeps each bilayer well aligned and prevents severe global disorder and frustration, while preserving the beating patterns (see Supplementary Information Fig. S1).

Unlike previous experimental studies focusing on symmetric alignment in the top and bottom heterostructures[37–39], the intricate interplay of high-symmetry stacking configurations in the R-H misaligned heterotrilayers unveils the elusive atomic registry textures. As depicted in Fig. 1d, an exemplary trilayer exhibits a top moiré with a 1° rotation (R-stack) and a bottom moiré with a 1.56° rotation (H-stack). Following the convention from Tong et al[36], these R- and H-stack moiré superlattices form a concatenated moiré pattern across the trilayer, delineated by nine distinct high-symmetry locales (Supplementary Information Fig. S2), each exhibiting different band hybridization between the top and bottom layers. At locales such as A$^R$/A$^H$ (similar for B$^R$/B$^H$, C$^R$/C$^H$ in Fig. S2), conservation of spin-valley degrees of freedom is anticipated to suppress tunneling and hybridization. By contrast, high-symmetry locales such as A$^R$/C$^H$ may allow stronger hybridization, depending on the local atomic alignment.

In reported heterobilayers, the spatial distributions of band-edge electrons differ between R-stack and H-stack heterostructures[40]. In R-stack heterobilayers, electrons are localized predominantly at the high-symmetry location C$^R$, whereas in H-stack heterobilayers, they are mostly found at location A$^H$. Combining R and H stackings in heterotrilayers likely generates multiple orbital configurations for electrons centered in the middle layer, where both A and C serve as trapping sites. The physics inherent in R-H style heterotrilayers promises a tapestry of intriguing phenomena.

**Multi-orbital nature of the bichromatic moiré superlattice**

The multi-orbital nature of the bichromatic moiré superlattice promotes the tunability of valley and spin degrees of freedom in TMD heterotrilayers. Experimentally, this manifests as multiple photoluminescence (PL) peaks in the heterotrilayer structure. The PL spectrum (Fig. 2a) reveals a dominant low-energy emission at ~1.39 eV and a secondary, weaker peak at ~1.46 eV. The energy separation of ~70 meV between the two peaks significantly exceeds the expected exciton double occupancy energy (~30 meV) at a single moiré site[26], indicating a multi-orbital origin. Power-dependent PL measurements on Device 3 (Fig. S3) show that both peaks persist even at the low excitation power, confirming their intrinsic origin from the heterotrilayer moiré potential rather than from power-induced filling effects. The normalized power dependence of the lower energy peak (Fig. S4) shows significant saturation behavior. The above evidence confirms the emissive excitonic species from heterotrilayer regions are intrinsic to the bichromatic moiré potential.

To probe the origins of the dual PL peaks, we conducted out-of-plane electric field measurements while keeping the doping fixed. Upon applying an electric field without external doping, both PL peaks exhibit clear linear Stark shifts (Fig. 2b). These linear Stark shifts correspond to the emissions originating from the interlayer excitons possessing significant out-of-plane dipoles.

The emission energy of the high-energy peak decreases monotonically with the applied electric field. This peak is attributed to the interlayer exciton, with its dipole moment oriented downward. The interlayer exciton arises from the binding of holes in the bottom $WSe_2$ layer and electrons in the middle $WS_2$ layer. Polarization analysis, shown in Fig. 2c, further supports the identification of this high-energy exciton. The peak exhibits cross-polarization ($\rho < 0$), indicating that the emission originates from an H-stack $WSe_2$-$WS_2$ heterobilayer[40]. This agrees with mid-bottom bilayer configuration.

In contrast, the low-energy peak shows opposite slopes when the sign of the electric field is reversed. These opposing slopes, observed in Fig. 2b-c, correspond to dipole moments that flip direction. This behavior confirms that the lower-energy exciton exhibits a bipolar emission characteristic. The dipole moment of this interlayer exciton (IX) flips dynamically with the applied electric field, with hole polarization occurring either in the top R-stacked layer or the bottom H-stacked layer (as shown in Fig. 2d-e).

At a small electric field near $E = 0.02$ V/nm, the low-energy interlayer excitons emitted from the top R-stacked and bottom H-stacked layers exhibit identical energies. In this range, the local slope of the low-energy peak is flat. This flat slope matches the quadrupolar excitons observed in symmetric heterotrilayers[37–39]. Despite differences in periodicity and stacking order between the top and bottom moiré superlattices, the identical energies and flat local slope are a signature of a shared ground state in the R-H aligned heterotrilayers. Additional data from another Device D2 is present in Supplementary Information Fig. S5.

The unique features of the PL peaks are attributed to the multi-orbital characteristics of the heterotrilayers. As illustrated in Fig. S2, the heterotrilayers feature multiple high-symmetry sites arising from the beating of the top and bottom moiré superlattices. Excitation of intralayer excitons in $WSe_2$ leads to rapid carrier relaxation to the band edges, forming interlayer excitons. At minimally hybridized sites, the excitons retain the same dipole moment. As a result, excitons emitted from these sites exhibit a single slope, as shown in Fig. 2b. By contrast, at strongly hybridized sites the lowest-energy state is shared by both the top and bottom layers. In this case, an applied electric field polarizes holes toward the bottom $WSe_2$ layer (for $E > 0$) or the top $WSe_2$ layer (for $E < 0$). Consequently, the bound excitons at these sites develop a positive or negative dipole moment depending on the direction of the electric field.

Although the excitons share the same lowest-energy at specific high-symmetry locations, the distinct stacking orders of the top and bottom bilayers enforce opposing selection rules. The low-energy PL peak for $E < 0$, characterized by a positive slope, exhibits a positive degree of polarization (Fig. 2c), which is consistent with the R-stack configuration in the upper heterobilayer of our device. In contrast, interlayer excitons in the lower heterobilayer, with negative slopes, display a negative degree of polarization, aligning with the H-stack configuration. This layer-polarized emission introduces an additional degree of freedom for manipulating circular polarization while maintaining fixed emission energies.

## Quadrupolar moiré trions

Under hole doping conditions, our experiments reveal the presence of quadrupolar excitonic emission in the moiré trilayers, as shown in Fig. 3a-b. Additional data across varying doping conditions from Device D1 are presented in Fig. S6. Furthermore, data from Device D2, tested under similar conditions, is included in Fig. S7. The formation of quadrupolar excitonic species occurs immediately upon introducing excess holes into the heterostructure. To confirm the formation of interlayer trions, we deliberately dope the heterotrilayer at fixed hole concentrations while sweeping the electric field. Fig. 3a and 3b show two hole-doping conditions at $n = -2.25 \times 10^{12}$ cm$^{-2}$ and $n = -3.75 \times 10^{12}$ cm$^{-2}$. The low-energy peak exhibits minimal shifts in emission energy under low-field conditions (additional hole doping scenarios are shown in Fig. S6). The flat slope confirms an almost negligible out-of-plane dipole moment. In contrast, the high-energy peak shifts negatively with the electric field, consistent with the behavior observed in Fig. 2b under intrinsic doping conditions. This shift is attributed to the fixed out-of-plane dipole moment of the exciton, whose position in the moiré superlattice results in a small hybridization probability, as illustrated in Fig. 2d-e.

Under electron doping conditions, the flat slope of the low-energy peak is absent (Fig. 3c). The observed slopes are directly ascribed to the dipoles associated with the top and bottom interlayer excitons.

We attribute the excitonic species with nearly zero out-of-plane dipole moments to the formation of quadrupolar moiré trions under hole-doped conditions. This interpretation is consistent with a scenario in which two holes are symmetrically distributed across the WSe$_2$ layers (Fig. 3a middle inset), regardless of whether the structure is R-stack or H-stack aligned. Strong Coulomb interactions drive the emergence of this unique three-particle complex, where two holes and one electron are strongly localized at the shared high-symmetry moiré sites. The resulting interlayer quadrupolar trion exhibits a nearly vanishing out-of-plane dipole moment, and its energy remains stable under varying electric fields, clearly distinguishing it from previously reported interlayer trions in heterobilayers[8,9,41]. Importantly, unlike the bosonic quadrupolar excitons observed in symmetric trilayers[37–39] that exist primarily under charge-neutral conditions, quadrupolar moiré trions appear only under doping. Furthermore, they require a substantially larger electric field to polarize into dipole configurations. This enhanced robustness indicates that the formation of interlayer quadrupolar moiré trions is energetically favored in bichromatic heterotrilayers.

When the electric field becomes sufficiently strong, holes in the bottom (top) WSe$_2$ layer are polarized to the top (bottom) WSe$_2$ layer to minimize the total energy, resulting in a well-defined out-of-plane dipole moment. Specifically, for hole density below $2 \times 10^{12}$ cm$^{-2}$ ($\sim v = -1$ for one moiré), a modest electric field of approximately $\pm 0.025$ V/nm is sufficient to transform the quadrupolar trion into a dipolar exciton (Fig. S6d-e). In this regime, polarized holes can redistribute across different moiré sites within the same WSe$_2$ layer, effectively lowering the total energy. However, at higher hole fillings $v \geq -1$, polarizing all holes into the same layer forces them to share moiré sites, incurring strong on-site Coulomb repulsion. Consequently, a larger field exceeding $\pm 0.05$ V/nm is needed to overcome this repulsion and achieve full dipolar polarization. These experimental findings demonstrate that the electric field can effectively tune the ground states of holes by manipulating their layer-specific degrees of freedom.

**Electric-field modulation of the moiré potential landscape**

We now deliberately tune the electric field to control the distribution of doped holes. As shown in Fig. 4a-b, at $E = 0$ V/nm, the shared ground state results in an even hole distribution across the top and bottom layers, with both WSe$_2$ layers experiencing the same moiré potential landscape. When the carrier density reaches $3.6 \times 10^{12}$ cm$^{-2}$, we observe a change in PL intensity (Fig. 4a) and enhanced polarization (Fig. 4b). This behavior aligns with the formation of an interlayer Mott insulating state, as reported by Lian et al[39]. However, in our study, this phenomenon occurs in an asymmetric heterotrilayer, characterized by distinct R-stack and H-stack layers, contrasting with the symmetric R-R stack with identical lattice constants in the top and bottom layers as described in previous work[39]. It's worth noting that, the interlayer Mott insulating state observed here differs from the behavior seen in regions with only bilayer R-stack heterostructure. In the R-stack heterobilayer, as shown in Fig. S8, holes are strictly confined to the top WSe$_2$ layer, and intensity variations occur at both fractional and integer filling factors.

When a fixed negative electric field (–0.05 V/nm) is applied, the doped holes would predominantly occupy the top WSe$_2$ layer. At a hole carrier density of $4\times10^{12}$ cm$^{-2}$, strong intensity modulation (Fig. 4c), and a significant decrease in polarization (Fig. 4d) are observed, as labelled by the dashed lines. The observed intensity and polarization modulations are consistent with previously reported signatures of correlated states in TMD heterobilayers[19,25–27,40,42]. It is worth noting that the emergence of integer-filling correlated states provides compelling evidence that the PL peaks originate from excitons localized at bichromatic moiré sites. The abrupt changes in PL intensity and polarization with doping, as shown in Fig. 4 and Figs. S8-10, are indicative of strong interactions between interlayer moiré excitons and correlated electronic states, consistent with prior observations in moiré systems[19,25–27,40,42].

By further tuning the electric field, the moiré potential profile is significantly altered, as shown in Fig. 4e-f. For a fixed electric field of –0.1 V/nm, we observe a surprising increase in the doping density required to reach the insulating state, rising to approximately $6.45\times10^{12}$ cm$^{-2}$. Furthermore, a distinct blueshift is observed at this point, indicative of the onset of on-site Coulomb repulsion. The substantial increase in carrier density needed to form the insulating state strongly suggests a shift in the density of ground-state sites.

Similar behavior is observed in Device D2 (Fig. S9). The electric field required to reshape the entire moiré landscape is slightly different, owing to the variation in twist angles between the R-stack and H-stack layers. As shown in Fig. S10, the moiré superlattice of the H-stack heterobilayer is approximately 7.6 nm, which is slightly larger than the value measured in Device D1 from the PFM data. However, qualitatively, the behavior with the applied electric field remains consistent with what we observed in Device D1.

To gain insight into the experimental observations, we have calculated the moiré potentials for valence band holes in heterotrilayer bichromatic superlattices, following the methodology outlined by Tong et al[36]. Detailed calculations are provided in the Supplementary Information. The concatenated moiré pattern is visualized by tracing several high-symmetry stacking sites, as shown in Supplementary Fig. S1, which includes both R-stack and H-stack high-symmetry regions. The calculated moiré potentials under various electric field conditions are presented in Fig. 5. The bright regions represent the local potential minima for holes (i.e., the valence band maximum). These regions will be first populated when the system is hole doped.

Considering the geometry of the trilayer moiré superlattice (Fig. 1d), we find that the heterotrilayer exhibits distinct moiré potential profiles compared to heterobilayers, with the added advantage of dynamic tunability via an applied electric field (Fig. 5). The simulations in Fig. 5 are performed under similar conditions to the measurements in Fig. 4. In Fig. 4a-b, the applied electric field is zero. At this condition, the hole density corresponding to the first integer filling insulating state, identified as an interlayer Mott insulator, is approximately $3.6\times10^{12}$ cm$^{-2}$, which is notably higher than the density required to reach $v = 1$ in the R-type heterobilayer with a moiré wavelength of 7.4 nm ($n \sim 2.1\times10^{12}$ cm$^{-2}$). In the absence of an electric field, the simulated moiré potential (Fig.

5a) shows bright regions slightly larger than the R-type stacking domains (Fig. S12), with multiple high-symmetry sites available for hole occupancy. As shown in Fig. 5b, the holes are distributed across both the upper and lower $WSe_2$ layers, forming a degenerate interlayer configuration. Laterally, the hole is confined to a common moiré potential minimum shared between layers, resulting in one hole occupying a shared ground state at the same high-symmetry moiré site in the hybridized trilayer. This correlated state is referred to as an interlayer Mott insulator[39]. In this regime, the holes can redistribute between the top and bottom $WSe_2$ layers within the same vertical moiré cells.

As the electric field increases ($E$-field pointing from the bottom gate to the top gate) to –0.05 V/nm, the hole density required for the Mott insulating state increases to $4.0 \times 10^{12}$ cm$^{-2}$ (Fig. 4c-d). The corresponding simulation shows the area of the bright regions expands slightly, indicating an enhanced capacity for hole occupancy (Fig. 5c). In this regime, holes preferentially occupy sites in the upper layer, with fewer holes residing in the lower layer (Fig. 5d). At an even higher electric field of –0.1 V/nm, the hole density at $v = -1$ further increase to $6.45 \times 10^{12}$ cm$^{-2}$ (Fig. 4e-f). The simulated potential landscape evolves into a honeycomb-like pattern (Fig. 5e), effectively doubling the number of sites available for hole localization, consistent with our experimental observations. In this regime, the electric field fully polarizes the doped holes to localize in the upper layer (Fig. 5f), where the correlated states are now governed by intralayer hole interactions with modified moiré periodicity and enhanced Coulomb repulsion. The strong on-site repulsion energy results in a significant blueshift in the excitonic emission. Overall, these results demonstrate that the applied electric field in a bichromatic moiré superlattice enables a tunable transition from interlayer- to intralayer-dominated Mott insulating states.

We realize that the role of strain relaxation in real devices would affect the moiré superlattice landscape. In complex trilayer structures, the R-stack and H-stack regions relax strain differently[40]. However, our calculations provide a qualitative physical framework for understanding the role of the electric field in modulating moiré potentials and hole distribution within the superlattices. These findings strongly support the realization of electrically tunable bichromatic moiré potentials.

## Discussion

In conclusion, we have demonstrated tunable excitonic and electronic states in uniquely engineered asymmetric R-H $WSe_2/WS_2/WSe_2$ bichromatic superlattices with two interfering moiré wavelengths. This configuration provides access to previously unexplored physical phenomena. First, the bichromatic superlattice hosts fermionic quadrupolar moiré trions, distinct from the bosonic quadrupolar excitons reported earlier. Second, both excitonic states and electronic ground states in these lattices can be highly tuned by external electric fields. Third, optical selection rules can be conveniently tailored within a single device. By carefully engineering stacking configurations in heterostructures, we demonstrate that quantum states in bichromatic moiré superlattices can be controlled and reconfigured, opening new pathways for investigating many-body physics in quantum materials.

Our findings open exciting possibilities for further investigation into the dynamics of multi-orbital excitons and other quasiparticles in more complex moiré superlattices. Future experiments would aim to explore the coevolution of multiple moiré potentials, particularly in trilayer systems with different stacking orders/angles, and their impact on charge/exciton interactions, and quantum phases. The ability to independently control multiple moiré superlattices holds the potential for realizing dynamically tunable quantum systems, with applications in quantum simulation, topological phases, and novel optoelectronic devices.

The emission of these quadrupolar excitons and trions involves a superposition of two excitonic species with dipole moments oriented in opposite directions. The investigation of the coherence and entanglement could provide new avenues for the manipulation of quantum light. Quadrupolar trions reported in this work are particularly robust against electrical noise under highly hole-doped conditions, which could emerge as promising stable sources.

## Methods

### Sample fabrication.

The sample fabrication details can be found in Ref[43]. In brief, a prefabricated bottom gate (a hexagonal boron nitride (hBN)/graphite stack) was transferred, and platinum contacts were defined above the stack using electron beam lithography. Subsequently, 3/7 nm of Cr/Pt was deposited via electron beam evaporation to form the bottom contacts. The bottom gate with contacts was then cleaned using an atomic force microscope (AFM) operated in contact mode.

Monolayer $WS_2$ and $WSe_2$ flakes were mechanically exfoliated from bulk crystals. The $WSe_2$ crystals were grown using the flux method, while the $WS_2$ crystals were sourced from HQ Graphene. Prior to stacking, the crystal orientations of the monolayers were determined via linear polarization-resolved second harmonic generation to ensure accurate alignment. To achieve the desired R- and H-stacking configurations, the top and bottom $WSe_2$ monolayers were taken from the same large flake, which was precisely cut using an AFM probe to minimize strain introduced by conventional tear-and-stack methods.

The monolayers were then assembled using a flat polycarbonate (PC)/polydimethylsiloxane (PDMS) stamp and transferred onto the cleaned bottom gate. Twist angles were confirmed using Piezoresponse Force Microscopy (PFM)[43,44]. PFM was carried out using a Bruker Dimension Icon AFM and Asylum Research Cypher S AFM with Pt–Ir-coated conductive probes (SCM-PIT-V2, force constant $\approx$ 3 N/m). An a.c. bias (<300 mV, $\approx$ 700 kHz and 350 kHz for lateral and vertical resonance frequencies, respectively) was applied between tip and sample, and the induced deformation amplitude and phase were recorded to probe the local electromechanical response. Once the moiré wavelength was determined, a top graphite/hBN stack was transferred for encapsulation and to complete the dual-gate geometry. Finally, Cr/Au electrodes for wire bonding were patterned using the electron beam lithography and deposited via e-beam evaporation (7/70 nm Cr/Au).

### Optical measurements.

PL measurements were performed using a home-built confocal microscope in reflection geometry. The sample was mounted in a close-cycled cryostat with temperature kept at 5 K (Montana CryoAdvance 50), unless otherwise specified. A helium-neon laser (1.96 eV) was used in PL measurements. An additional 650-nm short pass filter has been used to filter out any side bands. The excitation power for PL measurement was kept at 1 μW. The incident polarization was set to be $\sigma^+$ polarized using a linear polarizer and quarter-wave plate, while the photoluminescence was collected and polarization-resolved to $\sigma^+$ or $\sigma^-$ using another quarter-wave plate on a rotational mount and a linear polarizer. PL signals were dispersed by a diffraction grating and detected on a silicon charge-coupled device camera. The PL was spectrally filtered from the laser using a 700-nm long-pass filter before being directed into a spectrometer. A pinhole was used in the collection path to exclude signals from outside the laser spot area.

### Calculation of carrier density and electric field.

The carrier density is estimated using a parallel-plate capacitor model with the gate voltage applied. The BN thickness of samples was measured by AFM. D1 has 16 nm for top BN flake and 20 nm for the bottom BN flake. And the top and bottom BN thickness of D2 are ~27.5 nm. The carrier density is calculated using $C_t \Delta V_t + C_b \Delta V_b$, where $C_t$ ($C_b$) is the geometric capacitance of top (bottom) gate, and $\Delta V_t$ ($\Delta V_b$) is the effective doping voltage (the relative voltage from band edge). The value used for the dielectric constant of BN is $\varepsilon_{hBN} \approx 3$. The moiré wavelength $\lambda$ was derived directly from PFM images, and the corresponding moiré density is given as $n_0 \approx 2/(\sqrt{3}\,\lambda^2)$. The filling factor was calculated as $v = n/n_0$ and compared with the assignment of integer filling factors based on gate-dependent photoluminescence in well-understood heterobilayer region in the same sample.

The perpendicular electrical displacement field $D$ is calculated using $D = (C_t\,\Delta V_t - C_b\,\Delta V_b)/2\varepsilon_0$ and the electric field is calculated as $E = D/\varepsilon_{hBN}$.

**Theoretical calculations.**

The moiré potential energies are described in terms of the lowest several harmonics[36,45], wherein the parameters are determined by the high-symmetry stackings (See Supplementary Information for more details). All data for the high-symmetry stackings is extracted from first-principles calculations that are performed using the Vienna ab initio Simulation Package (VASP)[46].

## Data Availability

The Source Data underlying the figures of this study are available with the paper. All other datasets generated and/or analyzed during this study are available from the corresponding author upon reasonable request. Source data are provided with this paper.

## Acknowledgements

This work was mainly supported by the Wang Start-up funding funded by the College of Arts & Sciences at Washington University in St. Louis (WUSTL). The fabrication is partially supported by Ralph E. Powe Junior Faculty Enhancement Awards, partially supported by the Gordon and Betty Moore Foundation, grant DOI 10.37807/gbmf11560. X.W. acknowledges equipment support by the Center for Quantum Leaps at WUSTL. X.W. acknowledges the use of the Cypher S atomic force microscope for high-resolution PFM characterization of the moiré superlattices. The fabrication used instruments in the Institute of Materials Science and Engineering (IMSE) at WUSTL, with partial financial support from IMSE. Bulk $WSe_2$ crystals were grown and characterized by C.H., J.C. and J.Y. Materials synthesis by C.H. and J.C. was supported as part of Programmable Quantum Materials, an Energy Frontier Research Center funded by the U.S. DOE, Office of Science, BES, under award DE-SC0019443. J.Y. is supported by the US Department of Energy, Office of Science, Basic Energy Sciences, Materials Sciences and Engineering Division. K.W. and T.T. acknowledge support from the Elemental Strategy Initiative conducted by the MEXT, Japan (Grant Number JPMXP0112101001) and JSPS KAKENHI (Grant Numbers 19H05790, 20H00354 and 21H05233). C.Z. is supported by the Air Force Office of Scientific Research under Grant No. FA9550-20-1-0220 and the National Science Foundation under Grant No. PHY-2409943, OSI-2228725, ECCS-2411394. H.W. and L.Y. are supported by the National Science Foundation (NSF) grant No. DMR-2124934. The simulation used Anvil at Purdue University through allocation DMR100005 from the Advanced Cyberinfrastructure Coordination

Ecosystem: Services & Support (ACCESS) program, which is supported by National Science Foundation grants #2138259, #2138286, #2138307, #2137603, and #2138296. E. H. acknowledges support by the Gordon and Betty Moore Foundation, grant DOI 10.37807/gbmf11560.

## Author Contributions

X.W. conceived the project. X.W., M.C., R.L. fabricated the samples. X.W., M.C., R.L., Y.Y., Y.L. performed the measurements. X.W. analyzed and interpreted the results. H.W., L.Y. and C.Z. performed the calculations. T.T. and K.W. synthesized the hBN crystals. C.H., J.C., and J.Y. synthesized and characterized the bulk $WSe_2$ crystals. X.W., M.C., H.W., L.Y., C.Z., and E.H. wrote the paper with input from all authors. All authors discussed the results.

## Competing Interests

The authors declare no competing interests.

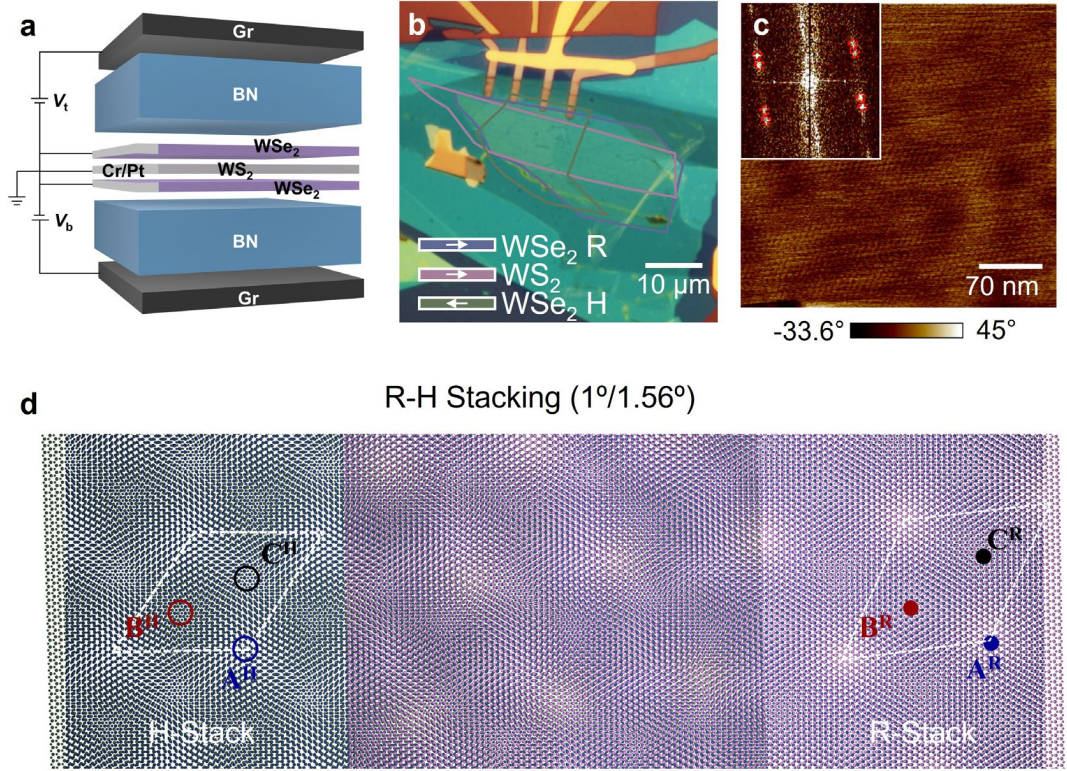

**Fig. 1. Structural characterization of trilayer heterostructure Device 1. a**, Schematic of a dual-gated WSe$_2$/WS$_2$/WSe$_2$ heterotrilayer device, in which pre-evaporated and AFM-cleaned Cr/Pt is used as contacts to the TMD layers. **b**, Optical image of Device D1 highlights the alignment of top WSe$_2$ and bottom WSe$_2$ layers relative to the middle WS$_2$ layer. Inset: Schematic illustrating layer alignments with arrows indicating R stacking (parallel) and H stacking (opposite). **c**, PFM image of Device D1's trilayer region, revealing two distinct moiré superlattices with lattice constants of approximately 7.4 nm (~0.9°) and 6.7 nm (~1.5°), circled in the inset. **d**, An exemplary trilayer with a top moiré with a 1° rotation (R-stack) and a bottom moiré with a 1.56° rotation (H-stack). The R- and H-stack moiré superlattices form a concatenated moiré pattern across the trilayer.

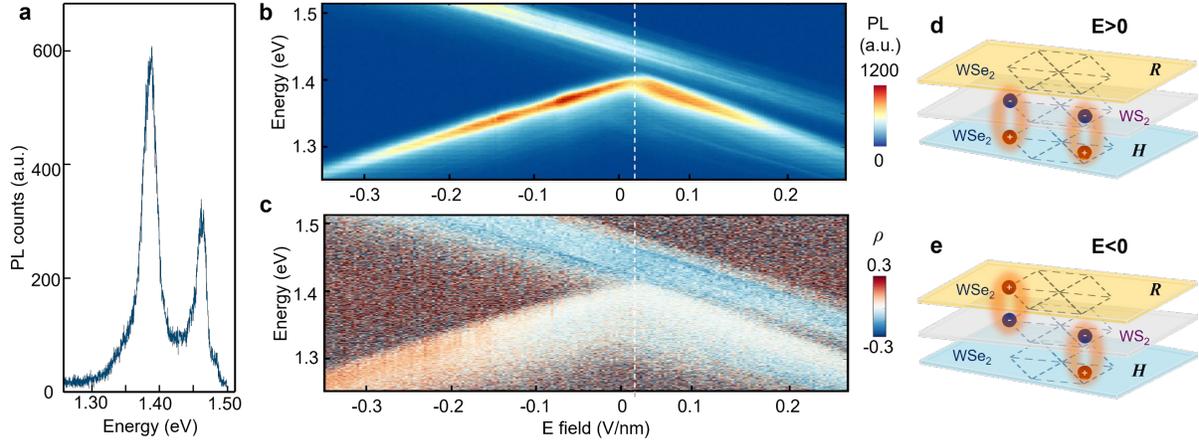

**Fig. 2. Electric field dependence of interlayer PL peaks. a**, PL spectrum of the trilayer heterostructure, displaying a dominant low-energy peak at 1.39 eV (low-energy peak) and a secondary peak at 1.46 eV (high-energy peak). **b**, Interlayer PL as a function of electric field without external doping possess different slopes of the Stark shift across positive and negative electric field regimes. **c**, The corresponding PL degree of circular polarization $\rho = \frac{\sigma^+/\sigma^+ - \sigma^+/\sigma^-}{\sigma^+/\sigma^+ + \sigma^+/\sigma^-}$ in the same conditions as in **b**. The high-energy peak remains cross-polarized regardless of field direction, whereas the low-energy peak flips from cross- to co-polarized when the electric field is reversed from positive to negative. **d-e**, Schematics of preferred sites of interlayer exciton emission with positive electric field (**d**) and negative electric field (**e**). The low-energy peak originates from sites where holes can redistribute freely between the top and bottom WSe₂ layers (the left side in **d**, **e**), whereas the high-energy peak comes from sites where holes remain localized in the bottom WSe₂ layer (the right side in **d**, **e**). In this study, positive electric field points from top WSe₂ to bottom WSe₂, and negative electric field points from bottom WSe₂ to top WSe₂ layer.

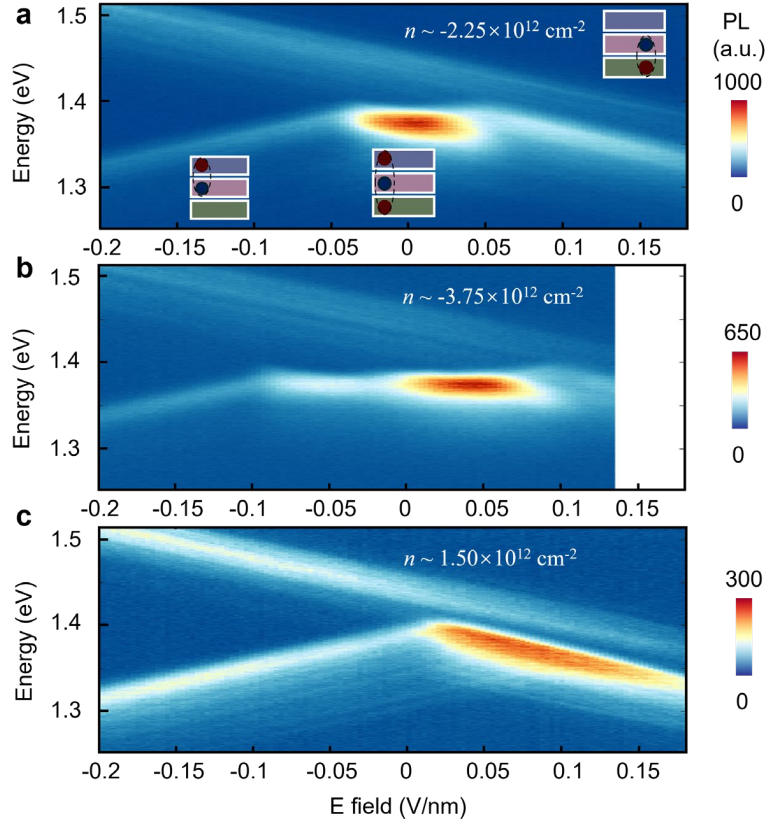

**Fig. 3. PL peaks' positions shift as a function of electric field under fixed doping conditions a,** $n = -2.25\times10^{12}$ cm$^{-2}$, **b,** $n = -3.75\times10^{12}$ cm$^{-2}$, **c,** $n = 1.5\times10^{12}$ cm$^{-2}$. Under hole doping (**a** and **b**), the peak energy of the low-energy peak remains nearly constant over a finite electric-field window, indicating the stabilization of a quadrupolar trion, whereas no such signature appears under electron doping (**c**). The inset of **a** schematically present, from right to left, the carrier configurations that yield the low-energy peak: holes localized in the bottom WSe$_2$ layer under positive electric field and pair with an electron in the middle WS$_2$ layer to form an interlayer exciton; holes shared between the top and bottom WSe$_2$ layers create a quadrupolar trion; and holes once again confined to the top WSe$_2$ layer bind to the middle-layer electron under negative electric field, forming another interlayer exciton. The blue and red circles represent electrons and holes located in the WS$_2$ and WSe$_2$ layers, respectively.

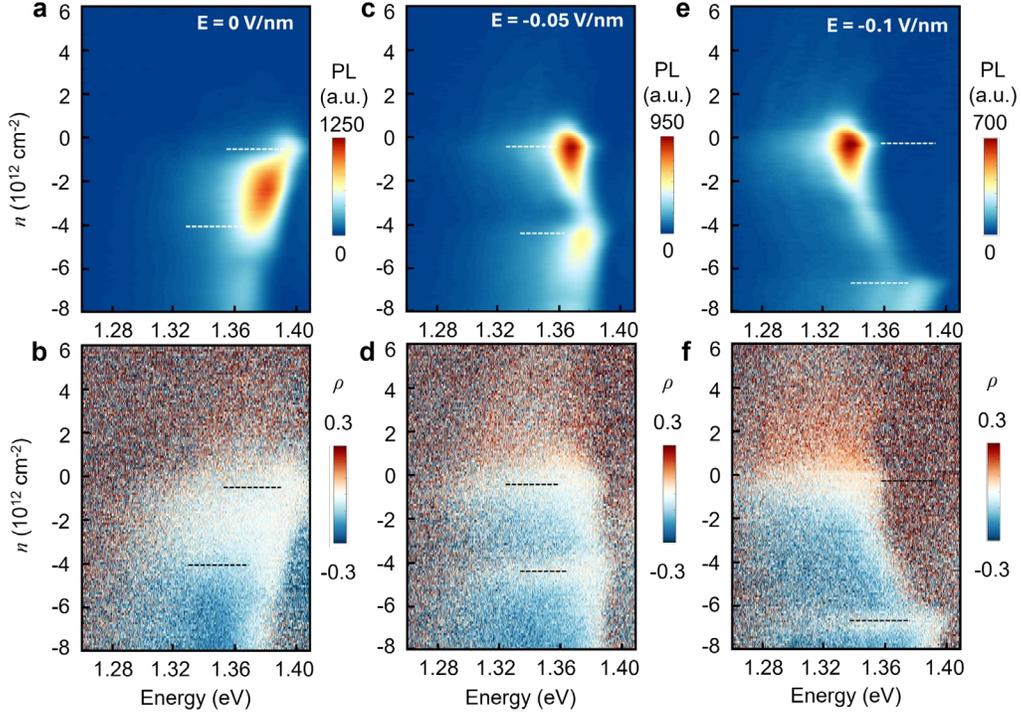

**Fig. 4. PL intensity and polarizationas a function of doping with fixed electric field. a-b**, PL intensity (**a**) and polarization (**b**) versus density with $E = 0$ V/nm. **c–d**, PL intensity (**c**) and polarization (**d**) versus density with $E = -0.05$ V/nm. **e–f**, PL intensity (**e**) and polarization (**f**) versus density with $E = -0.1$ V/nm. The PL intensity and polarization change are used to identify the integral filling and insulating states, $v = 0$ and $v = -1$ can be clearly discerned and marked by dashed lines in both the PL and corresponding polarization 2D maps. The corresponding density in $v = -1$ under the electric field of 0 V/nm, –0.05 V/nm and –0.1 V/nm is $3.6 \times 10^{12}$ cm$^{-2}$, $4 \times 10^{12}$ cm$^{-2}$ and $6.45 \times 10^{12}$ cm$^{-2}$, respectively. Line cuts of the raw data illustrating the polarization changes under specific conditions are provided in the Supplementary Information as Fig. S2.

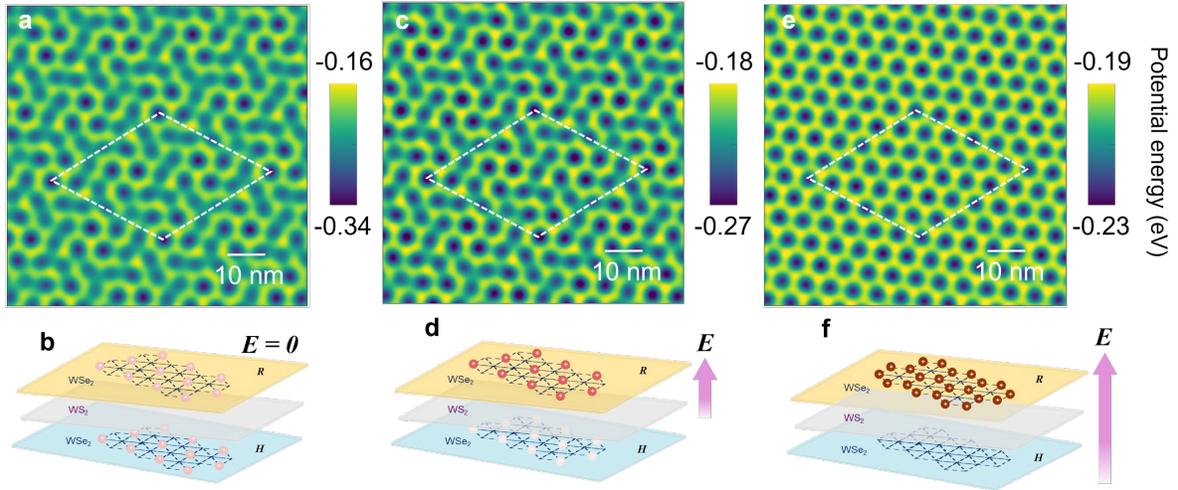

**Fig. 5. Moiré superlattice potential in the heterotrilayer without electric field (a), at electric field of −0.05 V/nm (c), and at electric field of −0.1 V/nm (e).** The bright yellow locales are potential minima for holes. **b, d, f,** Schematics showing the preferred layer localization of holes under the corresponding electric field conditions. As the out-of-plane electric field is swept from 0 to −0.05 V nm⁻¹ and then to −0.1 V nm⁻¹, the number hole-potential minima increase significantly, and the hole distribution shifts toward the top WSe₂ layer, accompanied by a transition of the Mott insulating state from interlayer to intralayer.

**Authors**: Mingfeng Chen[1,*], Runtong Li[1,*], Haonan Wang[1,*], Yuliang Yang[1], Yiyang Lai[1], Chaowei Hu[2], Takashi Taniguchi[3], Kenji Watanabe[4], Jiaqiang Yan[5], Jiun-Haw Chu[2], Erik Henriksen[1,6,7], Chuanwei Zhang[1,6,7], Li Yang[1,6,7], and Xi Wang[1,6,7,#]

[1]Department of Physics, Washington University, St. Louis, MO 63130, USA

[2]Department of Physics, University of Washington, Seattle, WA, USA

[3]Research Center for Materials Nanoarchitectonics, National Institute for Materials Science, 1-1 Namiki, Tsukuba 305-0044, Japan

[4]Research Center for Electronic and Optical Materials, National Institute for Materials Science, 1-1 Namiki, Tsukuba 305-0044, Japan

[5]Materials Science and Technology Division, Oak Ridge National Laboratory, Oak Ridge, Tennessee, 37831, USA

[6]Center for Quantum Leaps, Washington University, St. Louis, MO 63130, USA

[7]Institute of Materials Science and Engineering, Washington University, St. Louis, MO 63130, USA

**Table of Content**

**1. Moiré Wavelength Calculation in Heterotrilayer**

**2. High-Symmetry Locales in Moiré Heterotrilayer Superlattice**

**3. Power Dependent PL Intensity in Both Heterobilayer and Heterotrilayer Regions**

**4. Detailed Electric Field Dependent PL in Device 1 and Device 2**

**5. Doping Dependent PL in Device 1 and Device 2**

**6. Valley Polarizations Analysis**

**7. Moiré Potential Calculations**

**8. Density Functional Theory (DFT) Calculation Method**

# 1. Moiré Wavelength Calculation in Heterotrilayer

The commensurate moiré superlattice for the twisted trilayer $WSe_2/WS_2/WSe_2$ is formed by introducing relative twist angles between the top and bottom $WSe_2$ layers with respect to the middle $WS_2$ layer. This configuration generates a concatenated moiré superlattice, wherein the interference patterns of the individual heterobilayers combine to produce a larger, periodic moiré lattice. To find out the concatenated moiré superlattices, we first calculate moiré wavelength $\lambda$ of a heterobilayer $WSe_2/WS_2$ based on Ref. 1 and 2[1,2] with the following formula:

$$\lambda = \frac{(1+\delta)a_0}{\sqrt{2(1+\delta)(1-\cos\phi)+\delta^2}}, (1)$$

where $\delta = \left|\frac{a_{WSe_2}-a_{WS_2}}{a_{WS_2}}\right| = 4.1\%$ denotes the lattice mismatch between $WS_2$ and $WSe_2$. $a_0 = a_{WS_2} = 3.15$ Å is the lattice constant of monolayer $WS_2$, and the lattice constant of monolayer $WSe_2$ $a_{WSe_2} = 3.28$ Å. $\phi$ is the relative twist angle between two layers. We choose angles close to experimental parameters, with $\phi_R = 1°$, $\phi_H = 1.56°$. The corresponding moiré wavelengths are $\lambda_R = 7.3$ nm and $\lambda_H = 6.6$ nm, respectively. The moiré pattern orientation is then determined by

$$\tan\theta = \frac{-\sin\phi}{(1+\delta)-\cos\phi}, (2)$$

The relative twist angle between the two moiré bilayers is given by $|\theta_R - \theta_H|$. Substituting $|\theta_R - \theta_H|$ and $\left|\frac{\lambda_R-\lambda_H}{\lambda_H}\right|$ into eq. (1), the heterotrilayer moiré wavelength $\lambda$ for the twisted trilayer is computed as 33.6 nm. The commensurate condition for the twisted trilayer satisfies

$$\lambda_R\sqrt{m^2+n^2+mn} = \lambda_H\sqrt{p^2+q^2+pq}, (3)$$

where $m, n, p, q$ are integers. We find $\{m, n, p, q\} = \{5, -2, 6, -2\}$ that nearly satisfies eq. (3) and both hand sides in eq. (3) approach $\lambda = 33.6$ nm. The resulting commensurate moiré pattern is visualized by tracing the spatial arrangement of several high-symmetry stacking sites, including the R-type stacking sites ($A^R$, $B^R$, and $C^R$) and the H-type stacking sites ($A^H$, $B^H$, and $C^H$), as depicted in Fig. S1. These stacking sites reflect the local atomic registries of the trilayer structure, which are critical for determining the electronic and optical properties of the superlattice, as discussed in the following sections.

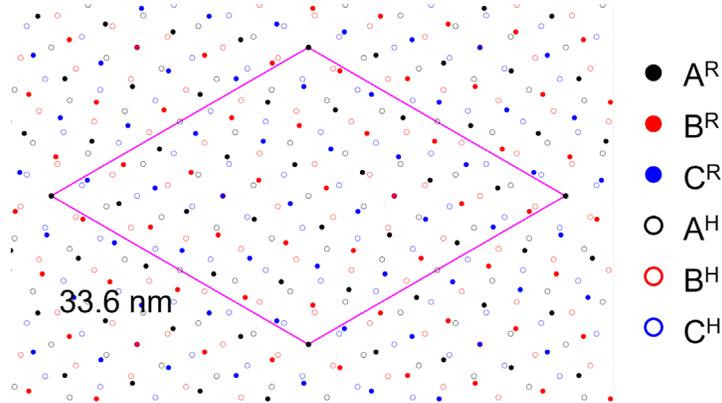

**Fig. S1** Commensurate moiré superlattices for twisted trilayer $WSe_2/WS_2/WSe_2$ by twisting the top and bottom $WSe_2$ with respect to the middle $WS_2$ layer by angles $\phi_R = 1°$, $\phi_H = 1.56°$. The concatenated moiré is visualized by tracing several high-symmetry stacking sites, including R-type stacking sites ($A^R$, $B^R$, and $C^R$) and the H-type stacking sites ($A^H$, $B^H$, and $C^H$).

## 2. High-Symmetry Locales in Moiré Heterotrilayer Superlattice

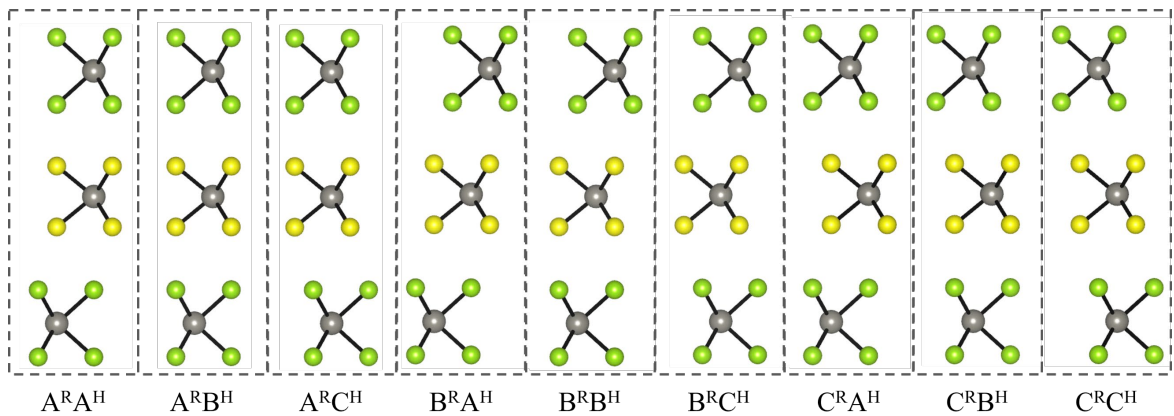

**Fig. S2.** Nine high-symmetry locales in R- and H-stack moiré heterotrilayer superlattice.

## 3. Power Dependent PL Intensity in Both Heterobilayer and Heterotrilayer Regions

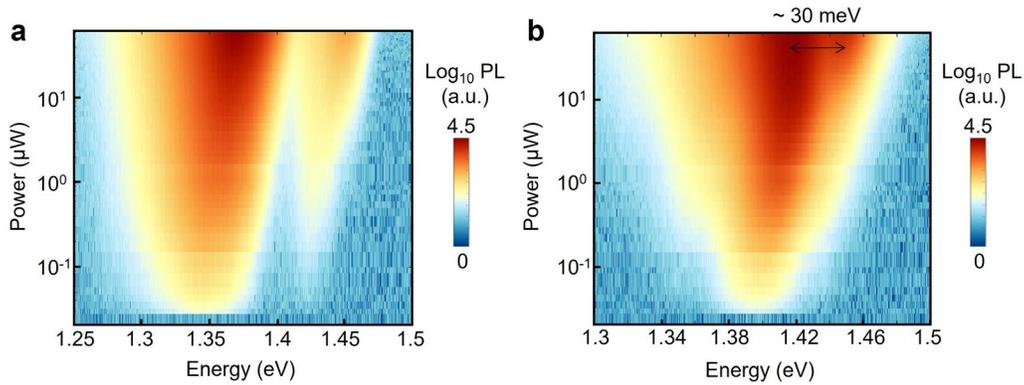

**Fig. S3. PL spectra at charge neutrality under circularly-polarized excitation and co-circular detection (σ⁺/σ⁺), in Device 3, as a function of excitation power in heterotrilayer (a) and heterobilayer (b) regions**. **a**, Power dependent PL in heterotrilayer region. Both the high-energy and low-energy peaks with energy difference ~70 meV are observed even with low excitation powers. **b**, Power dependent PL in bilayer region of the same sample, a different with a 30 meV higher energy starts to appear when the excitation power is above 20 μW. The PL intensity is plotted on a logarithmic scale to highlight weak spectral features and dynamic range.

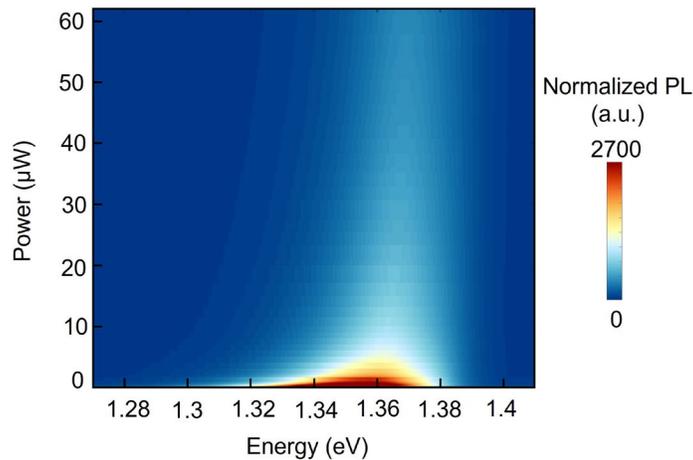

**Fig. S4. Normalized PL intensity as a function of excitation power.** The PL intensity is divided by the excitation power to highlight saturation behavior. The peak exhibits rapid saturation, suggesting the presence of a finite number of discrete localized states, which is consistent with excitons trapped in moiré potential minima.

## 4. Detailed Electric Field Dependent PL in Device 1 and Device 2

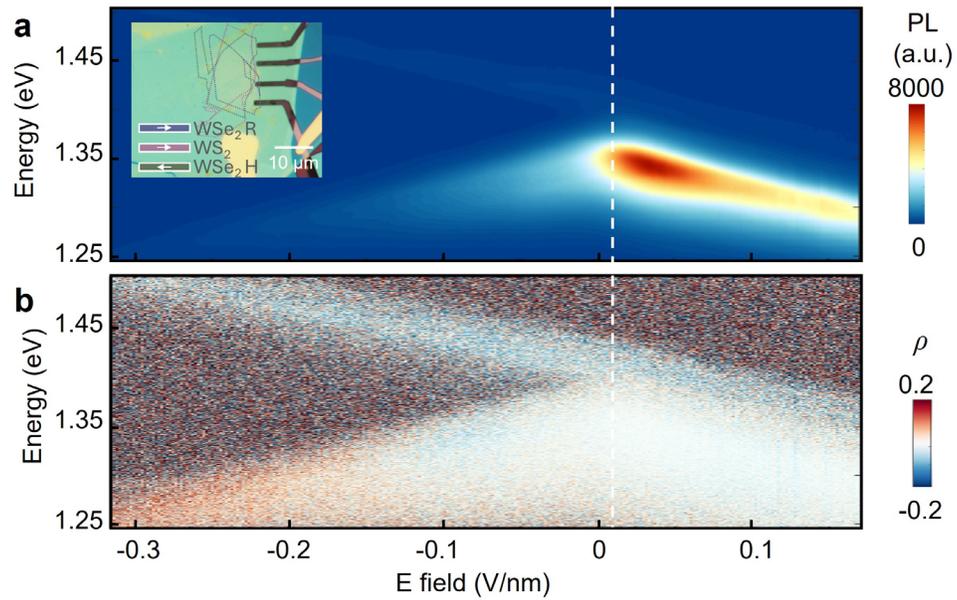

**Fig. S5. Electric field dependence of interlayer PL of heterotrilayer Device D2, without external doping. a,** Interlayer PL as a function of electric field. Inset: optical image and alignment configuration of the Device D2. **b,** The corresponding PL degree of circular polarization $\rho$.

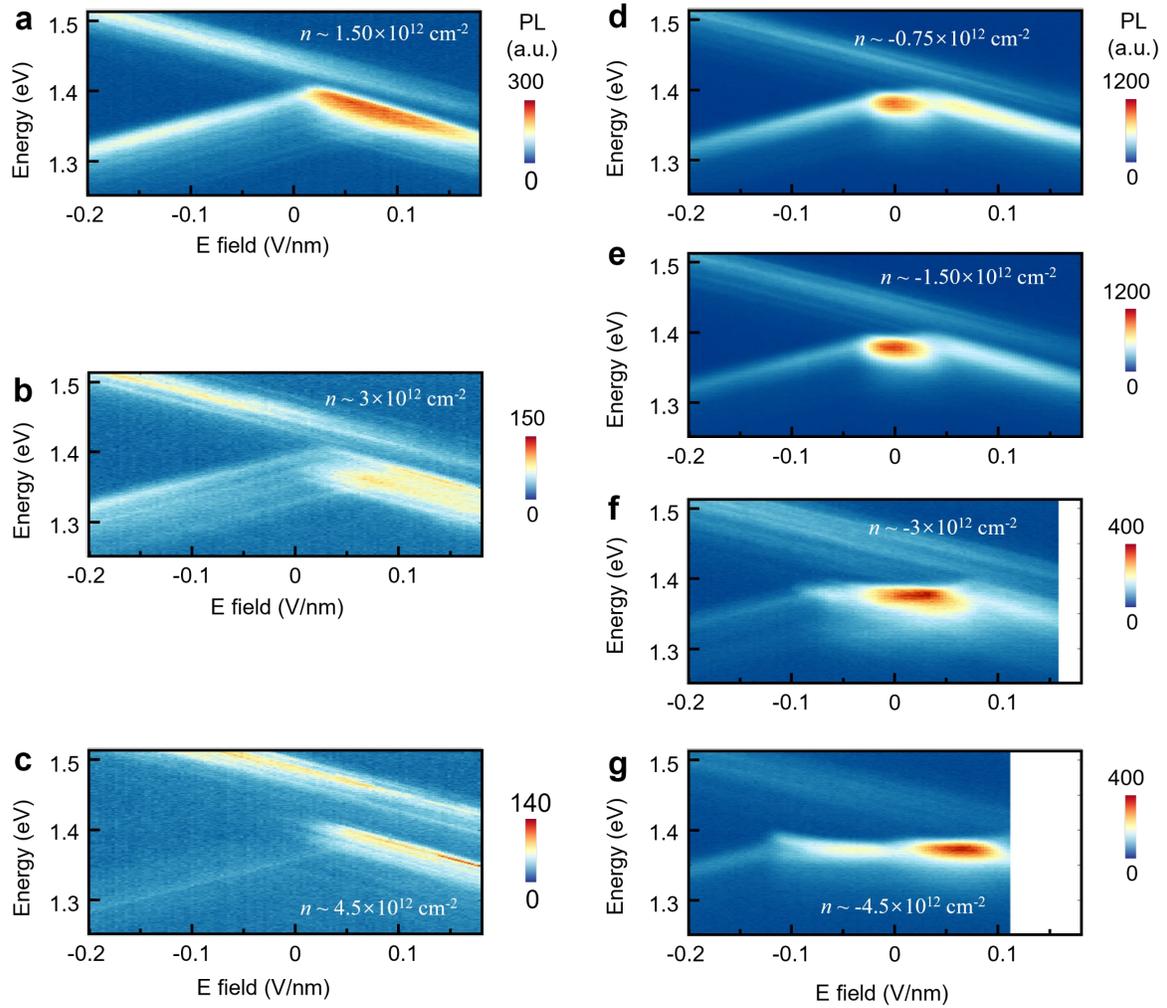

**Fig. S6. PL peaks' positions shift as a function of electric field at fixed doping conditions in Device D1. a–c,** At electron-doped conditions (**a**), $n = 1.5 \times 10^{12}$ cm$^{-2}$. (**b**), $n = 3 \times 10^{12}$ cm$^{-2}$. (**c**), $n = 4.5 \times 10^{12}$ cm$^{-2}$. **d–g,** At hole-doped conditions (**d**), $n = -0.75 \times 10^{12}$ cm$^{-2}$. (**e**), $n = -1.5 \times 10^{12}$ cm$^{-2}$. (**f**), $n = -3 \times 10^{12}$ cm$^{-2}$. (**g**), $n = -4.5 \times 10^{12}$ cm$^{-2}$.

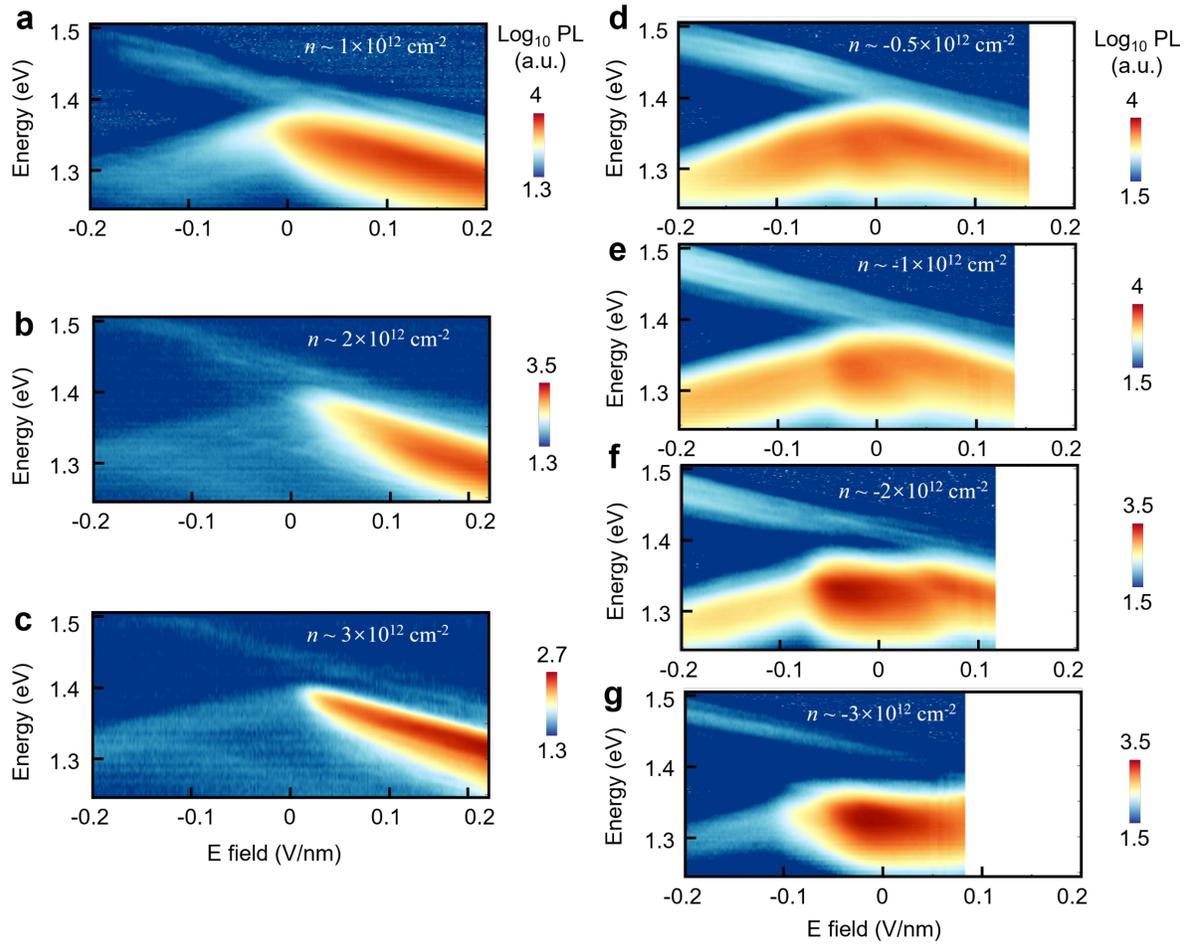

**Fig. S7. PL peaks' positions shift as a function of electric field under fixed doping conditions of heterotrilayer Device D2. a,** $n = 1 \times 10^{12}$ cm$^{-2}$. **b,** $n = 2 \times 10^{12}$ cm$^{-2}$. **c,** $n = 3 \times 10^{12}$ cm$^{-2}$. **d,** $n = -0.5 \times 10^{12}$ cm$^{-2}$. **e,** $n = -1 \times 10^{12}$ cm$^{-2}$. **f,** $n = -2 \times 10^{12}$ cm$^{-2}$. **g,** $n = -3 \times 10^{12}$ cm$^{-2}$.

## 5. Doping Dependent PL in Device 1 and Device 2

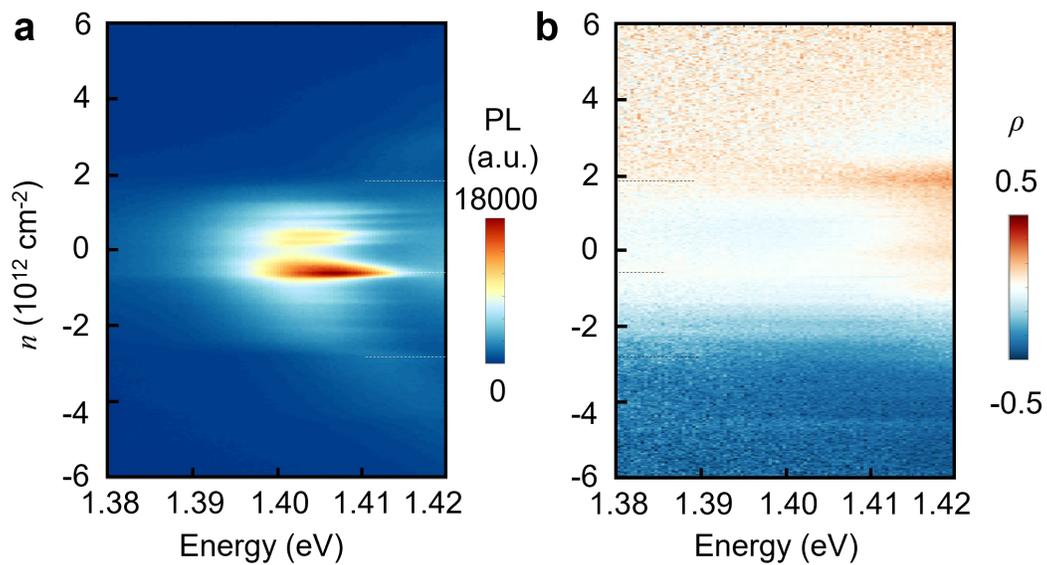

**Fig. S8. PL intensity (a) and polarization (b) as a function of doping from the R-stack heterobilayer region in Device D1.**

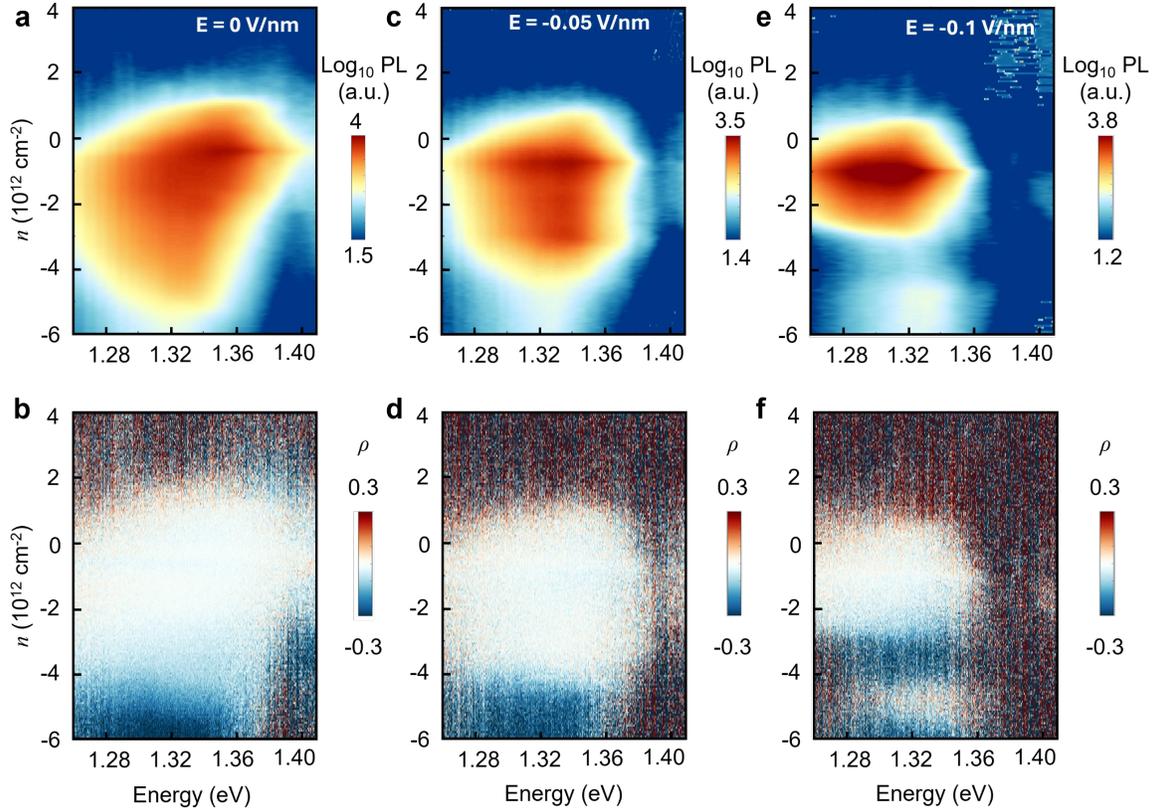

**Fig. S9. PL intensity and polarization as a function of doping with fixed electric field of trilayer heterostructure Device D2. a–b**, PL intensity (**a**) and polarization (**b**) versus density with $E = 0$ V/nm. **c–d**, PL intensity (**c**) and polarization (**d**) versus density with $E = -0.05$ V/nm. **e–f**, PL intensity (**e**) and polarization (**f**) versus density with $E = -0.1$ V/nm.

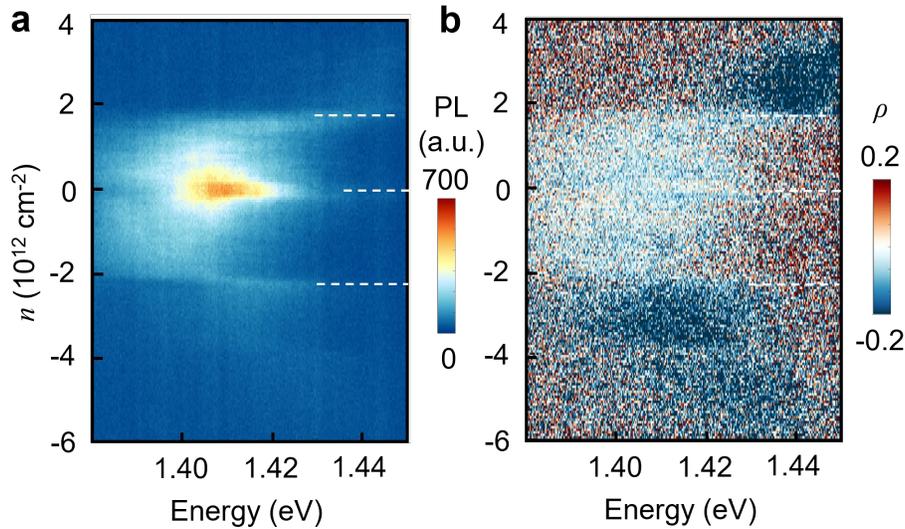

**Fig. S10. PL as a function of doping of the H-stack heterobilayer region in Device D2. a–b**, PL intensity (**a**) and polarization (**b**) versus doping.

## 6. Valley Polarizations Analysis

The excitation wavelength (633 nm) is off-resonance. As carriers relax to the band edge, multiple scattering processes lead to polarization loss. As a result, the absolute valley contrast is reduced but remains detectable.

To further substantiate the valley polarization and its evolution with doping, we display the raw co-polarized ($\sigma^+/\sigma^+$) and cross-polarized ($\sigma^+/\sigma^-$) spectra, together with two-dimensional maps of PL intensity and polarization versus carrier density for Device 1 (data also shown in Fig. 4 in the main text). Representative PL spectral lines collected under co- and cross-polarized detection are plotted for three characteristic regimes: charge neutrality $\nu = 0$, (yellow), $\nu = -1$ (red), and $\nu < -1$ (blue) in Fig. S11a-c and d-f; charge neutrality $\nu = 0$, (yellow), $\nu = -1$ (blue), and $\nu$ in between (red) in Fig. S11g-i.

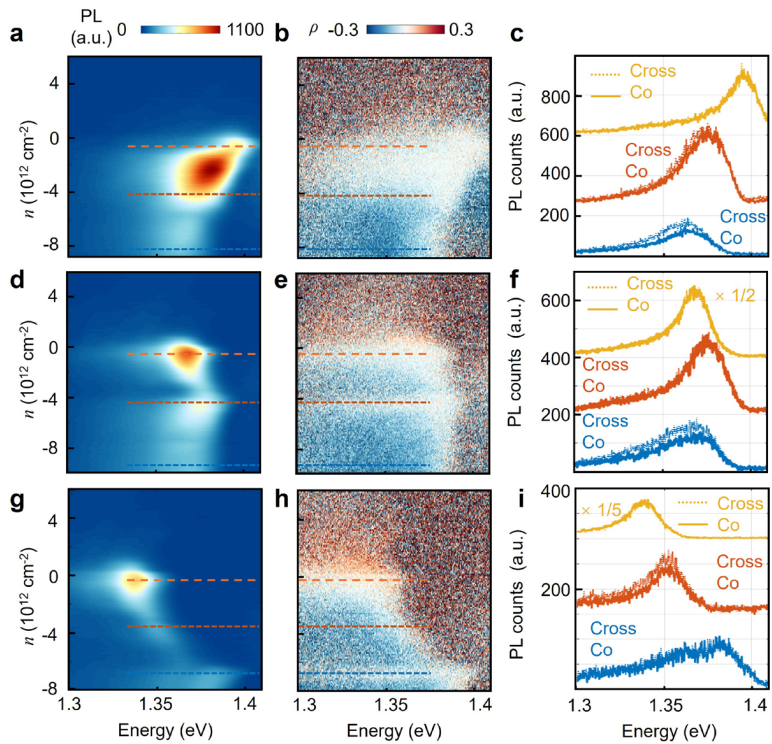

**Fig. S11. Doping- and polarization-resolved PL spectra under varying vertical electric fields.** The excitation wavelength (633 nm) is off-resonance. **a–c,** Measurements at zero vertical electric field. **a,** Doping-dependent PL intensity map. **b,** Polarization-resolved map. **c,** Co- (solid lines) and cross-polarized (dotted lines) PL spectra extracted from the doping levels indicated by dashed lines in **a** and **b**. **d–f,** Same set of measurements under an electric field of –0.05 V/nm. **g–i,** Same set of measurements under an electric field of –0.1 V/nm. In **f** and **i,** the top spectra are scaled by a factor of 1/2 and 1/5, respectively, for clarity. All spectra in **c, f, i** are vertically offset.

## 7. Moiré Potential Calculations

In this section, we describe the methodology for calculating the moiré potential, following the method published by Tong et al[3]. As discussed in the manuscript, the moiré transition metal dichalcogenide (TMD) trilayer system consists of two heterobilayers of $WSe_2/WS_2$ with distinct R-type and H-type stacking configurations. Due to the heterobilayer nature of these systems, out-of-plane electric polarization emerges spontaneously, resulting in a non-uniform charge distribution along the perpendicular axis[3]. Consequently, this charge asymmetry generates an electrostatic potential, referred to as the built-in voltage, across the TMD heterobilayers. Constrained by the three-fold rotational symmetry and lattice periodicity, $D(\mathbf{r})$ can be expanded by lowest several harmonics[3],

$$D(\mathbf{r}) = D_0 f_0(\mathbf{r}) + D_{+1} f_{+1}(\mathbf{r}) + D_{-1} f_{-1}(\mathbf{r}), \quad (4)$$

where $f_m(\mathbf{r}) = \frac{1}{9} \left| e^{-i\mathbf{K}\cdot\mathbf{r}} + e^{-i(\hat{C}_3\mathbf{K}\cdot\mathbf{r} - m\frac{2\pi}{3})} + e^{-i(\hat{C}_3^2\mathbf{K}\cdot\mathbf{r} + m\frac{2\pi}{3})} \right|^2$, with $\mathbf{K}$ being the wavevector at the corner of the monolayer Brillouin zone. The parameters $D_0$, $D_{+1}$ and $D_{-1}$ are determined from first-principles calculations for three high-symmetry stacking configurations ($A^R$, $B^R$, and $C^R$ for R-type stacking, and $A^H$, $B^H$, and $C^H$ for H-type stacking). For each stacking registry, the built-in voltage $D(\mathbf{r})$ is calculated as $\varphi_{WSe_2} - \varphi_{WS_2}$, representing the planar-averaged electrostatic potential difference across the TMD layers. The parameters are computed as $\{D_0^R, D_{+1}^R, D_{-1}^R\} = \{-152, -283, -120\}$ meV, and $\{D_0^H, D_{+1}^H, D_{-1}^H\} = \{202, 199, 398\}$ meV. The moiré potential for holes of the heterotrilayers is then given by $\frac{D_R(\mathbf{r}_R) - D_H(\mathbf{r}_H)}{2}$ [3]. Since the top (R-type stacking) and bottom (H-type stacking) have different periods, as discussed in the above section, the moiré superlattice potential has a larger wavelength (around 34 nm). When a perpendicular electric field is applied, a linear Stark shift[4] is induced to modify the built-in voltage. Due to the variation in interlayer distances across different stacking registries, the applied electric field results in distinct changes in the built-in voltage for each stacking configuration. For $E_g = -0.1 \, V/nm$, we find $\{D_0^R, D_{+1}^R, D_{-1}^R\} = \{-231, -222, -220\}$ meV, and $\{D_0^H, D_{+1}^H, D_{-1}^H\} = \{152, 160, 230\}$ meV.

The moiré superlattice potentials under different electric fields are depicted in Fig. S3. Within the moiré unit cell, several bright yellow regions are observed, representing potential minima for holes. When the moiré system is doped with holes, these regions are the first to be occupied. In the absence of an electric field, the area of the bright yellow regions is slightly more than that of the R-type stacking, several high-symmetry sites are allowed to be occupied by holes. As the electric field increases (applied from the bottom gate to the top gate) to –0.05 V/nm, the area of these regions slightly expands, indicating an increased capacity for hole occupancy. With a further increase in the electric field to –0.1 V/nm, the bright spots evolve into a honeycomb-like lattice structure, effectively doubling the number of holes. Although the upward electric field polarizes the holes toward the upper $WSe_2$ layer, it simultaneously expands the hole-potential minima to twice their original area. Consequently, the number of available trapping sites increases rather than decreases, so the hole density needed to reach $v = -1$ must likewise double, fully consistent with the experimental observations. These calculations of the moiré superlattice potentials provide a qualitative explanation for the experimental results, highlighting the role of the electric field in modulating moiré potentials as well as hole distribution within the superlattice.

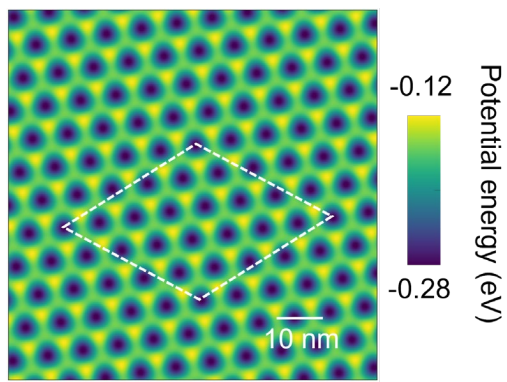

**Fig. S12. Moiré superlattice potential of R-stack heterobilayer.** The bright yellow dots are potential minima for doping holes.

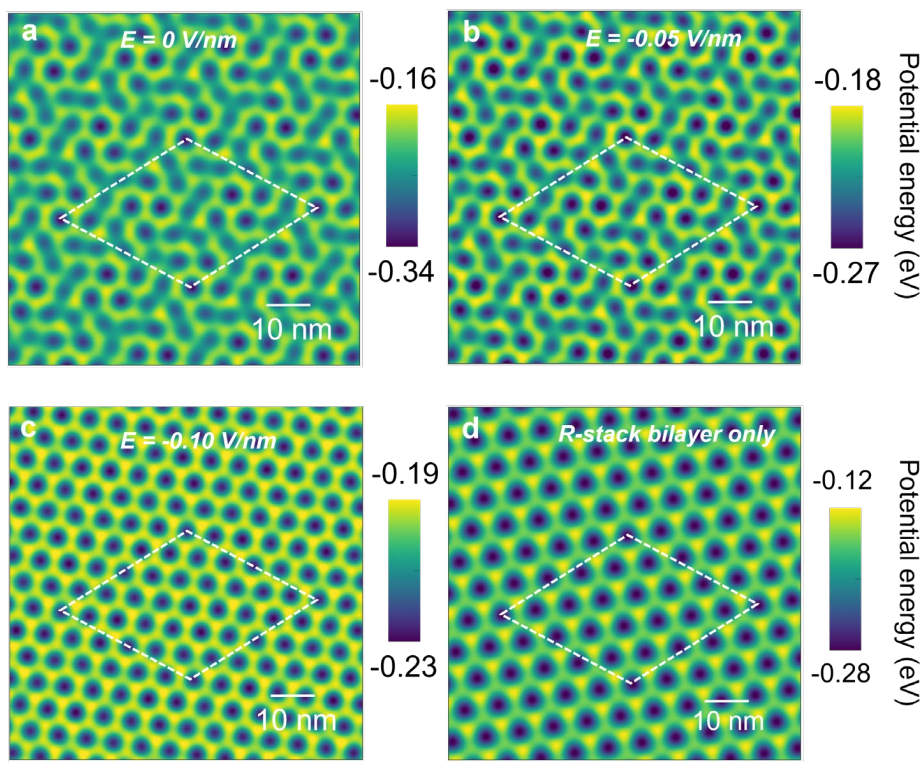

**Fig. S13** Moiré superlattice potential with electric field (**a**) 0 V/nm, (**b**) -0.05 V/nm, (**c**) -0.1 V/nm, and (**d**) bilayer R-type stacking. The bright yellow locales indicate potential minimum for doping holes.

## 8. Density Functional Theory (DFT) Calculation Method

The structural and electronic properties of the TMD heterobilayers are computed using density functional theory (DFT) as implemented in the Vienna Ab initio Simulation Package (VASP)[5,6]. The calculations employ the generalized gradient approximation (GGA) with the Perdew–Burke–Ernzerhof (PBE) exchange-correlation functional[7], utilizing projector-augmented wave (PAW) pseudopotentials[8] to treat core electrons. A Gamma-centered k-point grid of 15×15×1 is adopted for structural and electronic calculations, and a plane-wave energy cutoff of 450 eV ensures total energy convergence.

Spin-orbit coupling, which is critical for TMD heterobilayers, is incorporated throughout the calculations. Van der Waals interactions are accounted for using the DFT-D3 method with zero damping[9]. During structural relaxation, the in-plane lattice constants are fully relaxed, while the out-of-plane lattice constants are constrained. To mitigate long-range dipole interactions between polar slabs, a doubled unit cell with two symmetrically mirrored slabs is employed. A vacuum spacing of 15 Å is included to avoid spurious interlayer interactions.